\documentclass{elsart}

\usepackage{graphicx,natbib,amssymb}

\begin{document}

\begin{frontmatter}



\title{Enhancement of $T_{c}$ by Sr substitution for Ba in Hg-2212 superconductor}

\author[Grenoble]{P. Toulemonde\corauthref{cor}},
\corauth[cor]{Corresponding author. Present address: LPMCN,
Universit\'{e} Lyon-I, B\^{a}timent L\'{e}on Brillouin, 43
boulevard du 11 Novembre 1918, F-69622 Villeurbanne cedex,
France.} \ead{pierre.toulemonde@lpmcn.univ-lyon1.fr}
\author[Grenoble]{P. Odier}

\address[Grenoble]{Laboratoire de Cristallographie, CNRS, 25 avenue des martyrs, BP166, F-38042 Grenoble cedex 09, France.}


\begin{abstract}

The Ba substitution by Sr has been studied in two Hg-2212 series:
Hg$_{2}$(Ba$_{1-y}$Sr$_{y}$)$_{2}$YCu$_{2}$O$_{8-\delta}$ and
Hg$_{2}$(Ba$_{1-y}$Sr$_{y}$)$_{2}$(Y$_{0.80}$Ca$_{0.20}$)Cu$_{2}$O$_{8-\delta}$.
In both series a $T_{c}$ enhancement of about 40 K is observed
when Sr substitutes Ba from y~=~0 to y~=~1.0.

The y = 0 compound of the first series is the non superconducting
Hg$_{2}$Ba$_{2}$YCu$_{2}$O$_{8-\delta}$ prototype. In the second
series, this y = 0 compound is already superconducting at 21 K.
Indeed the members of this series present a higher charge carrier
density in their CuO$_{2}$ superconducting planes than their
homologues of the first series due to the doping introduced by the
substitution of 20~{\%} of Y by Ca. The compounds of both series
were synthesized in high pressure (3.5~GPa) - high temperature
(950~-~1050\r{ }C) conditions. In both cases Sr substitution was
successful up to the full replacement of Ba (y = 1.0). The Hg-2212
phases were characterized by XRD, SEM, EDX and a.c.
susceptibility.

\end{abstract}


\end{frontmatter}

\section{Introduction}

    The double-Hg-layer superconductor
Hg$_{2}$Ba$_{2}$YCu$_{2}$O$_{8-\delta}$ (Hg-2212) was discovered
in 1994 by Radaelli et al. \cite{Radaelli1}. This compound is an
insulator. When doped with Ca in
Hg$_{2}$Ba$_{2}$(Y$_{1-x}$Ca$_{x}$)Cu$_{2}$O$_{8-\delta}$, it
becomes superconducting \cite{Radaelli2,Radaelli3}. The T$_{c}$
transition depends on the Ca content. It increases from 40~K for
x~$\sim $~0.15 to 70~K for x~$\sim $~0.40. The optimal $T_{c}$ in
Hg-2212 system is 82-84 K \cite{Tokiwa1,Toulemonde1,Toulemonde2}.
This value can only be reached by chemical stabilization of the
oxygen-deficient double mercury layer. The oxygen vacancies
content in Hg$_{2}$Ba$_{2}$YCu$_{2}$O$_{8-\delta}$ is already
around ${\delta}$ = 0.4-0.5 and increases by Ca doping, up to its
solubility limit x~=~0.40. To further rise doping, stabilizing
elements like Tl$^{3+}$~\cite{Tokiwa2} or
Pb$^{4+}$~\cite{Toulemonde3,Toulemonde4} have to be used in
substitution of Hg$^{2+}$: these higher valency elements bring
oxygen in the double (Hg/Pb or Tl)$_2$O$_{2-\delta}$ layer and
then allow to further substitute Y by Ca.

The pressure effect on $T_{c}$ in Hg-2212 is huge, over + 50 K
under 20 GPa~\cite{Acha}. The corresponding rate is always
positive, independently of the doping state. Below 10~GPa, it
decreases from 4.5~K/GPa for underdoped compounds ($T_{c}\sim
$~30-45~K) to 2.7-3~K/GPa when $T_{c}\sim $~70~K. In optimally
doped (Hg,Pb)-2212 ($T_{c}\sim $~82-84~K) d$T_{c}$/dP~$\sim
$~1~K/GPa below 20~GPa~\cite{Toulemonde5}. By contrast, chemical
pressure, for example when Ba is replaced by a smaller cation (Sr
for instance), has a negative effect on
$T_{c}$~\cite{Wada,Subramian,Sin}. However, strontium substitution
does not reproduce the entire structural changes induced by
mechanical pressure in superconducting cuprates
\cite{Marezio1,Marezio2}. Two exceptions have been reported up to
now: the (La,Sr)$_{2}$CuO$_{4}$
system~\cite{Cava,Torrance,Radaelli4} and the artificially
stressed epitaxial (La,Sr)CuO$_{4}$ films \cite{Locquet}, where a
$T_{c}$ increase is observed induced by a chemical pressure.

In this paper, we have studied Ba substitution by Sr in two
Hg-2212 series of two different doping regimes (series I and II).
The first one, series I, corresponds to the composition
Hg$_{2}$(Ba$_{1-y}$Sr$_{y}$)$_{2}$YCu$_{2}$O$_{8-\delta}$ and is
built with Y in between the CuO$_{2}$ superconducting planes. The
second one, series II,
Hg$_{2}$(Ba$_{1-y}$Sr$_{y}$)$_{2}$(Y$_{0.80}$Ca$_{0.20}$)Cu$_{2}$O$_{8-\delta}$,
is doped with 20~{\%} of Ca on the Y site. Each sample was
synthesized at high pressure - high temperature (HP-HT) for
0~$\le$~y~$\le$~1.0. The effects of Sr/Ba replacement on
superconductivity were analyzed using X-ray diffraction (XRD),
scanning electron microscopy, EDX microanalysis and a.c.
susceptibility measurements.

\section{Experimental}

The samples were prepared in a high pressure Conac-type apparatus.
The reacting powders of each sample were mixed and compacted into
a gold capsule which was submitted to a high pressure (3.5~GPa) -
high temperature (1050\r{ }C for the Ca free series and 950\r{ }C
for the Ca doped series) treatment. At this point, it is important
to note that the chemical reactions take place into an inert and
closed capsule, i.e. into a confined system. Thus, the average
composition can not change during the HP-HT process, especially
the oxygen content of the sample, then the oxidation state of
copper in the final Hg-2212 compound is principally fixed by the
initial oxygen content of the starting mixture.

Two synthesis routes were tried, as explained in previous
papers~\cite{Toulemonde4,Toulemonde6}. The first one was used for
 series I (Ca-free). It consists in preparing two precursors,
``Ba$_{2}$YCu$_{2}$O$_{z}$'' and ``Sr$_{2}$YCu$_{2}$O$_{z}$'', by
reacting Y$_{2}$O$_{3}$ (Prolabo, 99.9~{\%}) and CuO (Aldrich,
99~{\%}) with BaO$_{2}$ (Merck, 95~{\%}) or SrCO$_{3}$ (Aldrich,
99.9~{\%}) at 850~\r{ }C or 950\r{ }C respectively during 24~h
under oxygen flow, followed by a quench to room temperature. Then,
this pre-reacted precursor (a mixture of Y$_{2}$BaCuO$_{5}$,
BaCuO$_{2}$, YBa$_{2}$Cu$_{3}$O$_{7-\delta}$ and
YBa$_{3}$Cu$_{2}$O$_{6.5+x}$ for the Ba-based precursor) is mixed
with HgO (Aldrich, $>$ 99~{\%}), inserted into the gold capsule
and treated at HP-HT. For
Hg$_{2}$(Ba$_{0.5}$Sr$_{0.5}$)$_{2}$YCu$_{2}$O$_{8-\delta}$, both
precursors were mixed together with HgO in the ratio 1:1:2.

In the second route we attempted to better control the oxidation
state of the sample by adjusting the oxygen content of the
precursor introduced into the closed capsule to the desired value.
This second method, used for series II, allows to control more
precisely the oxygen stoichiometry before the HP-HT reaction and
then the final doping state of the synthesized Hg-2212 compound.
It consists in intimately mixing HgO directly in the right
proportion with BaO$_{2}$, SrCuO$_{2}$, Y$_{2}$O$_{3}$,
Ca$_{2}$CuO$_{3}$ oxides and metallic Cu (Ventron, 99~{\%}) to
have the nominal
Hg$_{2}$(Ba$_{1-y}$Sr$_{y}$)$_{2}$(Y$_{0.80}$Ca$_{0.20}$)Cu$_{2}$O$_{7.50}$
composition, with y~=~0~-~0.25 by step of 0.05, then 0.40, 0.50,
0.70 and 1.0. For example, in the case of
Hg$_{2}$(Ba$_{0.75}$Sr$_{0.25}$)$_{2}$(Y$_{0.80}$Ca$_{0.20}$)Cu$_{2}$O$_{8-\delta}$,
to reach an oxygen content of 7.50, we used the mixture 2~HgO,
1.5~BaO$_{2}$, 0.5~SrCuO$_{2}$, 0.4~Y$_{2}$O$_{3}$,
0.1~Ca$_{2}$CuO$_{3}$ and 1.4~Cu. BaO$_{2}$ was chosen as the
barium source instead of BaCuO$_{2}$ because it is very difficult
to prepare BaCuO$_{2}$ without any trace of carbon contamination.
The 95~{\%} purity of the commercial BaO$_{2}$, which contains
traces of barium carbonate, was also not considered sufficient.
Our source of barium peroxide was prepared by precipitation of a
solution of barium nitrate (more pure, Aldrich, 99~{\%}) added
with oxygenated water. The absence of barium hydroxide, carbonate
or oxide hydrate was checked by XRD in the final BaO$_{2}$
product. SrCuO$_{2}$ was prepared by firing SrCO${_3}$ and CuO
under oxygen flow at 950~\r{ }C for 48~h followed by a second
treatment at 980~\r{ }C for 36~h with an intermediate grinding.
Ca$_{2}$CuO$_{3}$ was synthesized by reacting a mixture of
2~CaCO$_{3}$ (Aldrich, 99~{\%})~+~CuO at 1000~\r{ }C under oxygen
flow for 15~h. The purity of these two oxides was checked by XRD
before their mixing and the HP-HT treatment.

In the first set of syntheses, the obtained samples were
multiphasic with mixtures of Hg-2212, Hg-1212 and others
impurities~\cite{Loureiro1, Chaillout1}. In this work, the HP-HT
conditions (reaction time and temperature) were optimized in both
series to avoid the formation of the analogue Hg-1212 phase (which
appears for too long reaction time) and obtain the purest possible
Hg-2212 samples.

XRD patterns were collected using a powder Siemens D-5000
diffractometer working in transmission mode at the wavelength
$\lambda _{Cu,~K\alpha 1}$~=~1.54056~{\AA}. The microstructure and
composition of samples were investigated by scanning electron
microscopy on a JEOL-840, equipped with a Kevex system for X-ray
energy dispersive spectroscopy (EDX) analysis. A.C. susceptibility
measurements were performed at 119~Hz using a home-made apparatus
working at low magnetic field of 0.012~Oe, in the range
4.2~--~300~K (LEPES, CNRS, Grenoble, France).

\section{Results and discussion}

\subsection{\label{sec:level1}Phase purity and structural characterization}

\begin{figure}
\begin{center}
\includegraphics*[width=14cm]{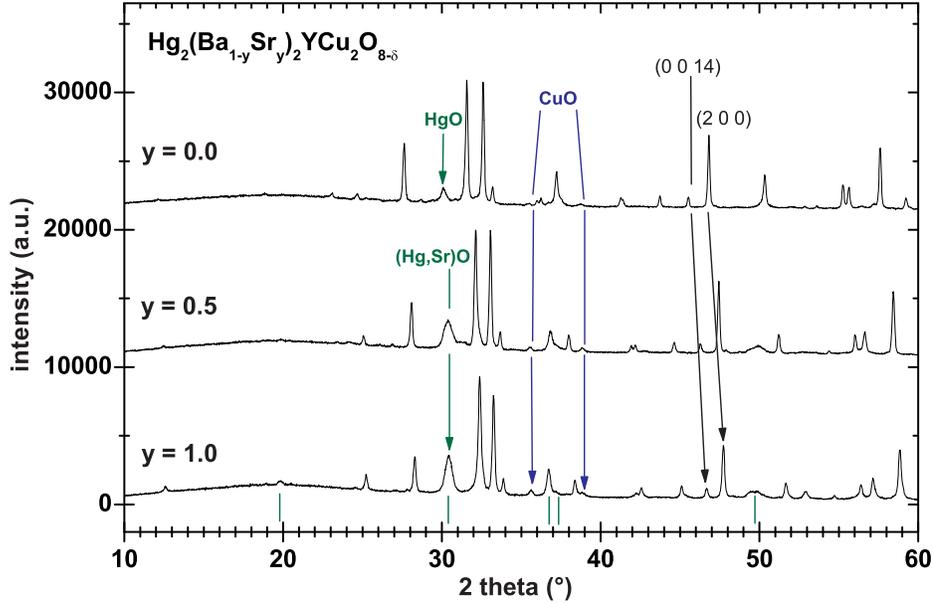}
\end{center}
\caption{X-ray diffraction patterns of the undoped Hg-2212 series
($\lambda _{Cu,~K\alpha 1}$~=~1.54056~{\AA}).}
\end{figure}

Figure 1 shows the XRD patterns of y~=~0, 0.5 and 1.0 compounds of
the series~I. Figure 2 shows a zoom of 20-35\r{ } and 45-60\r{ }
2${\theta}$ ranges for y~=~0, 0.20, 0.40 and 0.70 compositions of
the series II (x~=~0.20). In both series the main phase is
Hg-2212, without mono-Hg-layer Hg-1212 impurity. For the first
time Sr-substituted Hg-2212 compounds are synthesized at HP-HT
free of Hg-12(n-1)n type cuprates. Note also that full
substitution of Ba for Sr is reached in the Hg-2212 system.

\begin{figure}
\begin{center}
\includegraphics*[width=14cm]{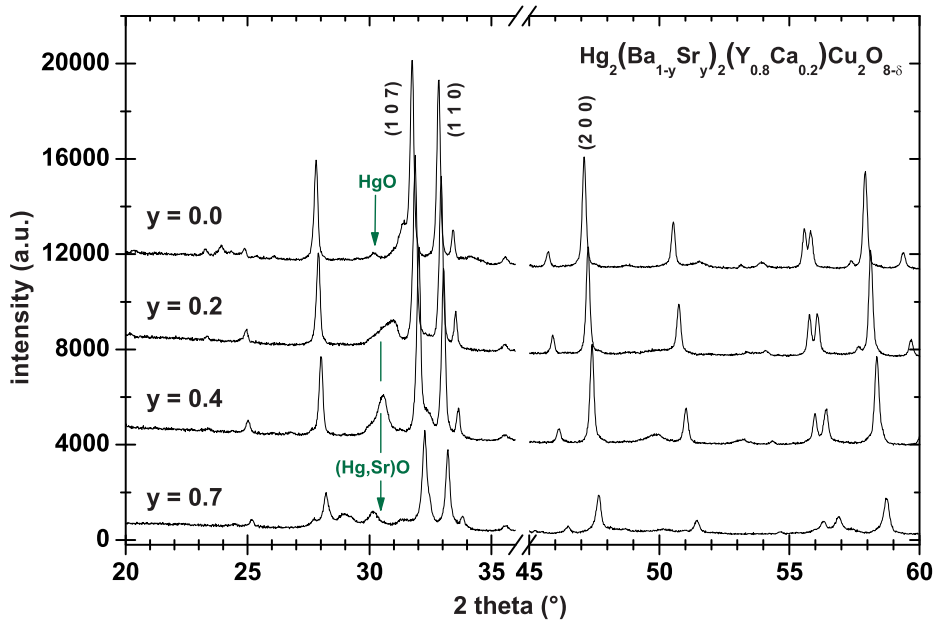}
\end{center}
\caption{X-ray diffraction patterns of the 20~{\%} Ca doped
Hg-2212 series.}
\end{figure}

For y~=~0.0, traces of HgO or CuO are observed. For y $>$ 0, an
unknown phase appears in both series, whose amount increases with
the Sr nominal content, it contains then Sr. The main XRD peak of
this phase is close to 31\r{ } 2${\theta}$ and its proportion
increases with the Sr content. EDX microanalyses made
systematically on samples of both series (SEM observations) show
that this phase contains mainly Hg and Sr. It could also
incorporate Cu or Ca in few amounts but the conclusion on this
point is not definitive. Nevertheless, the list of its main Bragg
peaks (indicated by tick marks in the bottom part of figure 1)
corresponds to that of the Ca-Hg-O phase observed as an impurity
in Ca-rich (Hg,Pb)-2212 compounds~\cite{Toulemonde3} and as an
intermediate phase in the Hg-1223 formation~\cite{Bordet1}. A
structural model (space group I~4/mmm) and a composition
(Ca$_{0.76}$Hg$_{1.24}$)O$_{2}$ were proposed for this new
phase~\cite{Bordet1}. This composition is close to that obtained
from our microanalysis where Sr replaces Ca:
(Sr$_{0.71}$Hg$_{1.29}$)O$_{2}$. These different observations
suggest a common structure for both phases, Ca- or Sr-based, with
the same stoichiometry.

\begin{figure}
\begin{center}
\includegraphics*[width=10cm]{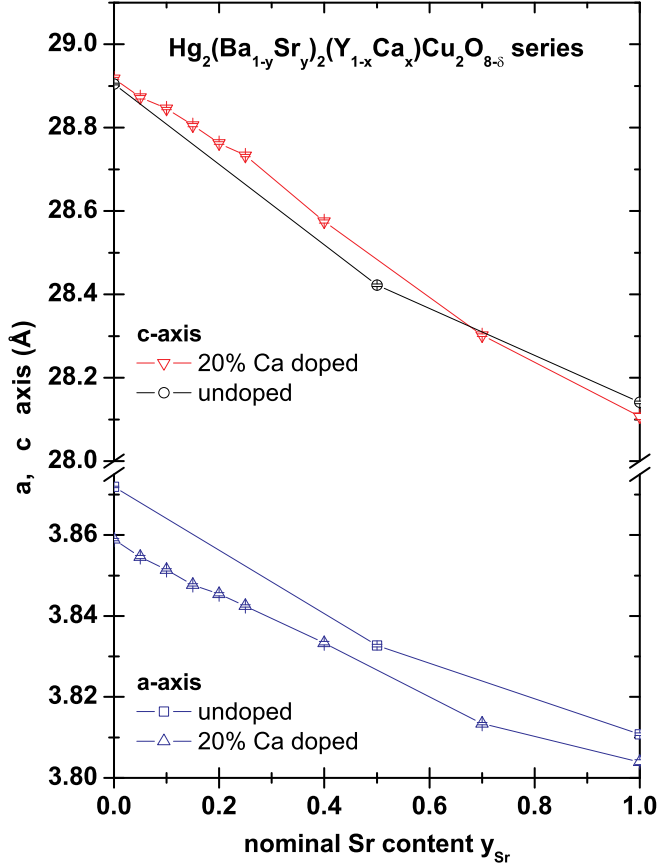}
\end{center}
\caption{Lattice parameters versus nominal Sr content in both
Hg-2212 series (x~=~0 and 0.20).}
\end{figure}

\begin{table}
\caption{\label{tab:table1} Lattice parameters and $T_{c}$ for
Hg$_{2}$(Ba$_{1-y}$Sr$_{y}$)$_{2}$YCu$_{2}$O$_{8-\delta}$ samples
(series~I).}
\begin{tabular}{lccc}
\hline
 Sr content &y~=~0.0 &y~=~0.5 &y~=~1.0\\ \hline
 a = b ({\AA})& 3.8719(2)& 3.8327(3)& 3.8108(3) \\
 c ({\AA})& 28.905(2)& 28.422(3)& 28.141(3) \\
 $T_{c}$ onset& 0~K& 32~K& 42~K \\ \hline
\end{tabular}
\end{table}

The shift of the peaks towards higher 2${\theta}$ values shows the
shrinkage of the lattice due to the smaller Sr. It is particulary
obvious on figure 2 for (107), (110) and (200) reflections. It
proves that the Sr substitution on Ba site is effective. EDX
microanalyses also show the Sr substitution and suggest that it
reaches the nominal content, i.e. the measured Sr/(Sr+Ba) ratio is
close to the nominal Sr stoichiometry. From EDX, the Ca doping is
under-stoichiometric, in the 15-20~{\%} range with respect to
nominal stoichiometry (20~{\%}). Lattice parameters were
calculated by least square method (tables 1 and 2). In both series
a decrease of the lattice parameters is observed, with a similar
rate, i.e. $\sim $~1.4-1.6~{\%} along the a-axis and $\sim
$~2.6-2.8~{\%} along the c-axis, as shown on figure 3. The lattice
contraction is then not completely isotropic, it is larger for
c-axis due to a larger compressibility along this direction than
in the basal plane (i.e. the CuO$_{2}$ superconducting plane)
which contains rather rigid bonds of O-Cu-O type. The shrinkage of
the a-axis is similar to that observed for the equivalent axis in
Y-123 or Y-124 compounds~\cite{Licci,Karpinski}. The c-axis
compression is around 0.76-0.82~{\AA} for full substitution. This
value is only slightly smaller than that predicted on considering
the replacement of Ba by Sr in the four Ba-O planes of the
structure and the difference of ionic radius between Ba and Sr (in
coordination number 9). The full substitution would give a
contraction of 1.28~{\AA}. The a-axis for the two fully
Sr-substituted samples is around 3.80-3.81~{\AA}, which is typical
of superconducting compounds containing only Sr, like Bi-based
cuprates~\cite{Hazen1,Tarascon1}. Moreover, no accident is visible
on the variation of a-axis (and c-axis) which could indicate a
brutal change of stoichiometry, on the oxygen composition for
example. We can conclude that the variations of a and c lattice
parameters are exclusively due to the Sr substitution for Ba.

\begin{table}
\caption{\label{tab:table2} Lattice parameters and $T_{c}$ for
Hg$_{2}$(Ba$_{1-y}$Sr$_{y}$)$_{2}$(Y$_{0.80}$Ca$_{0.20}$)Cu$_{2}$O$_{8-\delta}$
samples (series~II).}
\begin{tabular}{lccccccccc}
\hline
Sr content &y~=~0.0 &y~=~0.05 &y~=~0.1 &y~=~0.15 &y~=~0.20 &y~=~0.25 &y~=~0.40 &y~=~0.70 &y~=~1.0 \\
 a = b ({\AA})& 3.8589(3)& 3.8546(4)& 3.8514(3)& 3.8477(3)& 3.8454(4)& 3.8425(3)& 3.8333(4)& 3.8134(3)& 3.8039(6) \\
 c ({\AA})& 28.918(3)& 28.873(4)& 28.846(3)& 28.806(4)& 28.763(3)& 28.733(3)& 28.575(4)& 28.302(3)& 28.107(6) \\
 $T_{c}$ onset& 21~K& 27~K& 32.5~K& 39.5~K& 38.5~K& 41~K& 51~K& 60~K& 58~K \\ \hline
\end{tabular}
\end{table}

\subsection{\label{sec:level2}Comparison of chemical pressure and mechanical pressure}

\begin{figure}
\begin{center}
\includegraphics*[width=10cm]{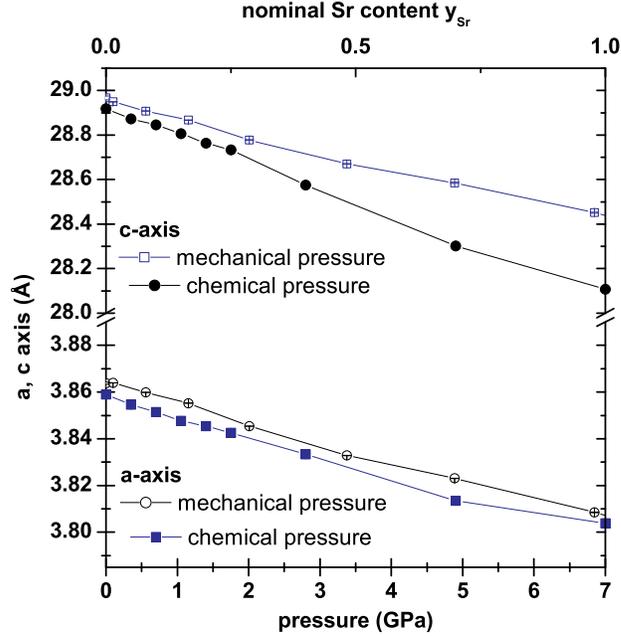}
\end{center}
\caption{Lattice parameters versus nominal Sr content for the
Hg-2212 Ca-doped series (this work) and versus mechanical pressure
for Hg$_{2}$Ba$_{2}$(Y$_{0.85}$Ca$_{0.15}$)Cu$_{2}$O$_{8-\delta}$
(ESRF ID-09 data, P.~Bordet et al.~\cite{Bordet2},
S.~M.~Loureiro~\cite{Loureiro1}).} \label{fig:exmp1}
\end{figure}

Chemical pressure (by Sr) and mechanical pressure effects on
lattice parameters are compared in figure 4. These data are
reproduced from a previous high pressure experiment performed in a
diamond anvil cell on the ID-09 beamline at the european
synchrotron ESRF (Grenoble, France) with a
Hg$_{2}$Ba$_{2}$(Y$_{0.85}$Ca$_{0.15}$)Cu$_{2}$O$_{8-\delta}$
compound~\cite{Loureiro1,Bordet2}. The scale of both X-coordinates
(nominal Sr content and mechanical pressure) was chosen to have
approximatively the same decrease for a-axis in both cases.

We observe on the c-axis variations that the effect of mechanical
pressure is not fully comparable with that of Sr chemical
pressure. The pressure necessary to obtain the same shrinkage of
the lattice corresponding to the full substitution Sr/Ba is not
the same if one considers a-axis or c-axis. Along the a-axis it
requires the application of an equivalent pressure of~$\sim
$~7~GPa whereas this value is higher along the c-axis, $\sim
$~10~GPa. The difference is larger than experimental errors. In
Y-123 and Y-124 systems, the complete substitution of Ba with Sr
would induce a variation of the lattice parameters equivalent to
the application of an external pressure of about
10-9.3~GPa~\cite{Licci,Karpinski}. This extrapolated value is
consistent with what is observed here for the Hg-2212 system.

According to Acha et al.~\cite{Acha}, a pressure of 10~GPa should
increase $T_{c}$ by 30-45~K, depending on the doping state of the
Hg-2212 compound. We will compare this value to the increase of
$T_{c}$ obtained by chemical pressure of Sr in both series in the
next paragraph.

\subsection{\label{sec:level3}Superconducting properties}

\begin{figure}
\begin{center}
\includegraphics*[width=10cm]{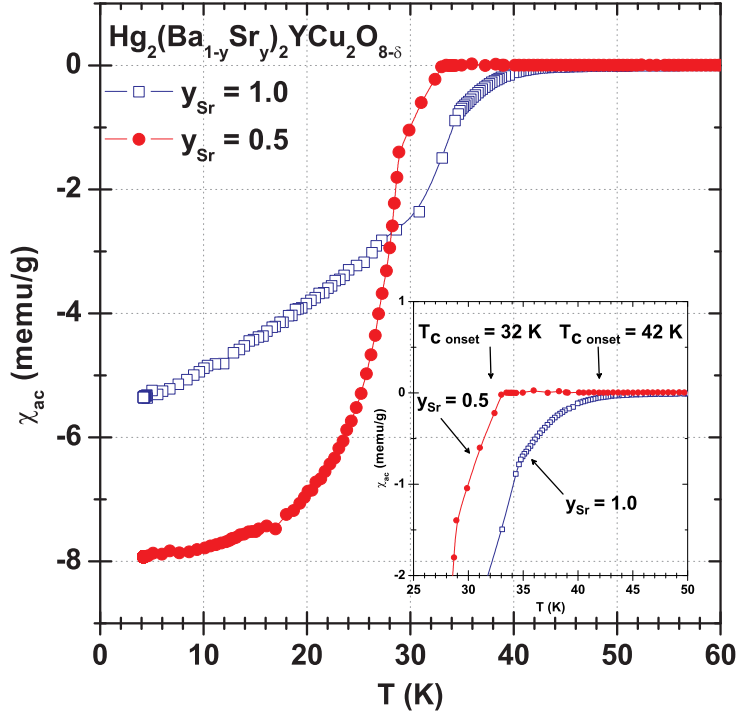}
\end{center}
\caption{A.C. susceptibility versus temperature for y = 0.50 and
1.0 compounds of series~I.} \label{fig:exmp2}
\end{figure}

\begin{figure}
\begin{center}
\includegraphics*[width=12cm]{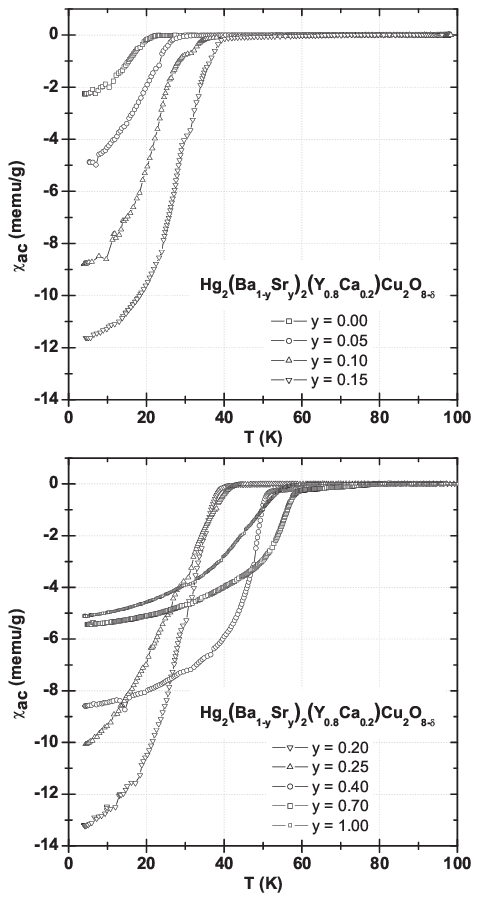}
\end{center}
\caption{A.C. susceptibility versus temperature for y = 0.0-1.0
compounds of series~II.} \label{fig:exmp3}
\end{figure}

The critical temperature $T_{c}$ was determined by a.c.
susceptibility on fine powders which is a rather good method to
determine the superconducting volume fraction (at 119~Hz with a
low field of 0.012~Oe). Figures 5 and 6 show the measurements for
series~I and II. All the samples were superconducting, except the
one for y~=~0.0. The observed superconducting volume fractions (up
to 70~{\%}) do not correspond to the entire volume of the samples
but are sufficient to attribute the transition to Hg-2212 which is
the only cuprate detected by XRD.

The first compound of series I,
Hg$_{2}$Ba$_{2}$YCu$_{2}$O$_{8-\delta}$, was not superconducting.
This result is in agreement with previous measurements made by
Radaelli et al. who did not observed superconductivity for this
composition~\cite{Radaelli1,Radaelli3}. They showed that Ca doping
on the Y site was necessary to induce superconductivity in
Hg-2212. We show here, for the first time, that Ca doping on Y
site is not necessary to reach the superconducting state. Indeed
Hg$_{2}$(Ba$_{0.5}$Sr$_{0.5}$)$_{2}$YCu$_{2}$O$_{8-\delta}$ and
Hg$_{2}$Sr$_{2}$YCu$_{2}$O$_{8-\delta}$ are already
superconducting without Ca doping, respectively at 32~K and 42~K
(table 1). As a result of series~I, $T_{c~onset}$ increases
regularly from 0~K (y~=~0) to 42~K (y~=~1)  with the Sr content.
In series~II, the
Hg$_{2}$Ba$_{2}$(Y$_{0.80}$Ca$_{0.20}$)Cu$_{2}$O$_{8-\delta}$
compound, is superconducting at 21~K. This value is consistent
with those reported previously. As for the previous case
(table~2), the Ba replacement by Sr in series~II also increases
$T_{c}$ continuously with the Sr content from 21~K (y~=~0) to
60-58~K (y~=~0.7-1.0).

To summarize, a 40~K increase is observed in
Hg$_{2}$(Ba$_{1-y}$Sr$_{y}$)$_{2}$(Y$_{1-x}$Ca$_{x}$)Cu$_{2}$O$_{8-\delta}$
(0~$\le$~y~$\le$~1.0; x~=~0, x~=~0.2), contrarily to other
Sr-substituted families (Y-123 or Hg-12(n-1)n) where a decrease of
$T_{c}$ is shown~\cite{Wada,Subramian,Sin}. This is a remarkable
case where the Sr-chemical pressure has a positive effect on
$T_{c}$. This increase of 40~K is in the same order of magnitude
than that obtained by applying a mechanical pressure of about
10~GPa in the Ba-based Hg-2212 compound~\cite{Acha}. In addition,
the rate of $T_{c}$ enhancement with Sr/Ba substitution is similar
in both series, as illustrated in figure 7. As a consequence, it
does not seem to be dependent of the doping level. Nevertheless,
is the doping level constant among both series, i.e. equal to the
doping regime of the y~=~0 compound~? This point is discussed in
the next paragraph.

\subsection{\label{sec:level4}Discussion}

Both EDX and lattice parameters variations agree to show that Sr
substitutes to Ba nearly at the nominal content. A first
consequence is that Y$^{3+}$ is not substituted by Sr$^{2+}$. This
is discussed more carefully below.

If Sr$^{2+}$ would substitute Y$^{3+}$ on its site, then it would
affect the hole doping of the CuO$_{2}$ planes. This does not
happen for different reasons. First, the EDX results show that Sr
and Ba contents vary in opposite way: the Sr content increases
regularly when the Ba content decreases, and more precisely, the
ratio of contents Sr/(Sr+Ba) determined by EDX is close to the
nominal Sr content in all samples. Secondly, the $T_{c}$ rise is
regular and continuous. If Sr$^{2+}$/Y$^{3+}$ site substitution
would have been involved in this $T_{c}$ rise, this would have
resulted in an over-stoichiometry of Sr. Moreover, this
over-stoichiometry should have increased regularly to explain the
$T_{c}$ increase. None of those phenomena is observed. Thirdly, if
one assumes that Sr$^{2+}$ goes on the Y$^{3+}$ site and that its
occupancy factor increases, for steric considerations one would
observe an increase of the c-axis (the ionic radius of Sr$^{2+}$
being larger than Y$^{3+}$ in the eight-fold coordination), or at
least a levelling of the shrinkage of the c-axis. This is neither
observed.

As a consequence, because the substitution of Ba$^{2+}$ to
Sr$^{2+}$ is isoelectronic, the doping is not changed in both
series when the Sr content varies from y~=~0 to 1.0. Probably the
oxygen content in the Hg-2212 phase does not change significantly
neither. This point is clearer in the series~II, where the oxygen
control was more accurate (mixture of oxides and metallic Cu with
a total oxygen content of O 7.50) than in series~I (mixture of HgO
with precursors whose oxygen content is not well defined).

If the doping state is not changed and if the origin of this
$T_{c}$ enhancement is structural, other experiments are necessary
to precise its structural origin. We have performed neutron powder
diffraction on both series to know exactly the changes induced
locally into the 2212 structure. Refinements are underway and the
corresponding results will be published
elsewhere~\cite{Toulemonde7}. One significant result concerns the
oxygen content determined from Rietveld refinements: it is almost
constant in the whole range of Sr substitution for
Ba~\cite{Toulemonde5}. It means that the $T_{c}$ enhancement
observed in both series is not related to hypothetical oxygen
content changes, but more likely a consequence of the Ba
replacement by Sr.

\begin{figure}
\begin{center}
\includegraphics*[width=10cm]{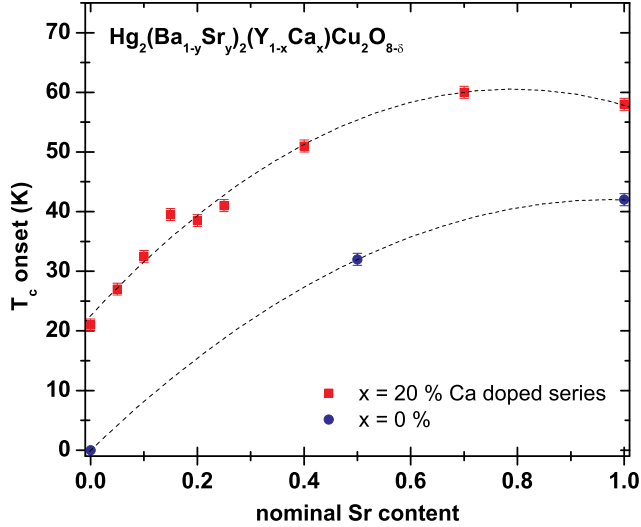}
\end{center}
\caption{Critical temperature versus nominal Sr content in both
Hg-2212 series (x~=~0 and 0.20).}
\end{figure}

\section{Conclusion}

The effects of chemical pressure were investigated in two Hg-2212
series: one based on yttrium,
Hg$_{2}$(Ba$_{1-y}$Sr$_{y}$)$_{2}$YCu$_{2}$O$_{8-\delta}$
(series~I), and the second one doped with 20~{\%} of Ca,
Hg$_{2}$(Ba$_{1-y}$Sr$_{y}$)$_{2}$(Y$_{0.80}$Ca$_{0.20}$)Cu$_{2}$O$_{8-\delta}$
(series~II). For the first time, Sr-substituted Hg-2212
superconductors were synthesized at high pressure~-~high
temperature free from other analogue cuprate phases, i.e. members
of the mono-Hg-layer family Hg-12(n-1)n.

We have shown that doping is not required to induce
superconductivity in Hg-2212 system:
Hg$_{2}$(Ba$_{0.5}$Sr$_{0.5}$)$_{2}$YCu$_{2}$O$_{8-\delta}$ and
the full Sr-substituted Hg$_{2}$Sr$_{2}$YCu$_{2}$O$_{8-\delta}$
compounds of series~I are already superconducting without Ca
doping. In series~II the Sr substitution enhances
superconductivity that is already present in the Sr-free
Hg$_{2}$Ba$_{2}$(Y$_{0.80}$Ca$_{0.20}$)Cu$_{2}$O$_{8-\delta}$
compound.

In both series $T_{c}$ increases continuously from 0~K (y~=~0) to
42~K (y~=~1.0) for the first series and from 21~K to 60~K in the
second one. In conclusion, contrarily to most superconducting
cuprates, the Sr-chemical pressure has a positive effect on
$T_{c}$ in the Hg-2212 system.




{\bf Acknowledgements}

P. Toulemonde thanks CNRS for financial support during his PhD
research. The authors are grateful to Dr J.~L.~Tholence (LEPES,
CNRS, Grenoble, France) for his help in the a.c. susceptibility
measurements, to Dr~P.~Bordet for providing us the ESRF high
pressure diffraction data, and to Dr~S.~Le~Floch for her help in
the use of the Conac press. The authors acknowledge also R.
Argoud, C. Brachet and R.~Bruy\`{e}re for technical assistance in
HP-HT experiments.


\begin{thebibliography}{}


\bibitem{Radaelli1} P. G. Radaelli, M. Marezio, M. Perroux, S. de Brion, J.L.
Tholence, Q.~Huang, and A. Santoro, Science 265 (1994) 380.
\bibitem{Radaelli2} P. G. Radaelli, M. Marezio, J. L. Tholence, S. de Brion, S.
Loureiro, A. Santoro, Q.~Huang, J.J. Capponi, M. Alario-Franco,
and C. Chaillout, Physica~C 235-240 (1994) 925.
\bibitem{Radaelli3} P. G. Radaelli, M. Marezio, J. L. Tholence, S. d. Brion, A.
Santoro, Q. Huang, J. J. Capponi, C. Chaillout, T. Krekels, and G.
V. Tendeloo, Journal of Physics and Chemistry of Solids 56 (1995)
1471.
\bibitem{Tokiwa1} A. Tokiwa-Yamamoto, T. Tatsuki, S. Adachi, and K. Tanabe, Physica C 268 (1996) 191.
\bibitem{Toulemonde1} P. Toulemonde, P. Bordet, J. J. Capponi, J. L. Tholence, and P.
Odier, Inst. Phys. Conf. Ser. No 167 Applied Superconductivity
(1999) 271.
\bibitem{Toulemonde2} P. Toulemonde, A. Sin, P. Bordet, J. L. Tholence, and P. Odier,
Physica~C 341-348 (2000) 677.
\bibitem{Tokiwa2} A. Tokiwa-Yamamoto, T. Tatsuki, X.-J. Wu, S. Adachi, and K.
Tanabe, Physica C 259 (1996) 36.
\bibitem{Toulemonde3} P. Toulemonde, P. Odier, P. Bordet, R. Bruy\`{e}re, and J.L.
Tholence, Physica C 366 (2002) 147.
\bibitem{Toulemonde4} P. Toulemonde, P. Odier, P. Bordet, C. Brachet, and E. Suard,
Physica C 377 (2002) 146.
\bibitem{Acha} C. Acha, S. M. Loureiro, C. Chaillout, J. L. Tholence, J. J.
Capponi, M.~Marezio, and M. Nunez-Regueiro, Solid State
Communications 102 (1997) 1.
\bibitem{Toulemonde5} P. Toulemonde, Ph.D. Thesis, Grenoble, France, October 2000.
\bibitem{Wada} T. Wada, S. Adachi, T. Mihara, and R. Inaba, Jpn. J. Appl.
Phys., Part 2 26 (1987) L706.
\bibitem{Subramian} M. A. Subramian, and M. H. Whangbo, J. Solid State Chem. 109 (1993) 410.
\bibitem{Sin} A. Sin, F. Alsina, N. Mestres, A. Sulpice, P. Odier, and M.
N\'{u}\~{n}ez-Regureiro, J. Solid State Chem. 161 (2001) 355.
\bibitem{Marezio1} M. Marezio and F. Licci, Physica C 282-287 (1997) 53.
\bibitem{Marezio2} M. Marezio and F. Licci, Supercond. Sci. Technol. 13 (2000) 451.
\bibitem{Cava} R. J. Cava, R. B. van Dover, B. Batlogg, and E. A. Rietman,
Phys. Rev. Lett. 58 (1987) 408.
\bibitem{Torrance} J. B. Torrance, Y. Tokura, A. I. Nazzal, A. Bezinge, T. C.
Huang, and S.~S.~P. Parkin, Phys. Rev. Lett. 61 (1988) 1127.
\bibitem{Radaelli4} P. G. Radaelli, D. G. Hinks, A. W. Mitchell, B. A. Hunter, J.
L. Wagner, B. Dabrowski, K. G. Vandervoort, H. K. Viswanathan, and
J. D. Jorgensen, Phys. Rev. B 49 (1994) 4163.
\bibitem{Locquet} J.-P. Locquet, J. Perret, J. Fompeyrine, E. M\"{a}chler, J.
W. Seo, and G.~Van Tendeloo, Nature 394 (1998) 453.
\bibitem{Loureiro1} S. M. Loureiro, PhD thesis, 1997, Grenoble, France.
\bibitem{Chaillout1} C. Chaillout, S.M. Loureiro, P. Toulemonde, J.J. Capponi, and M.~Marezio,
Physica C 282-287 (1997) 895.
\bibitem{Toulemonde6} P. Toulemonde, S. Le~Floch, P. Bordet, J.J. Capponi, M.~M\'{e}zouar, and P. Odier,
Supercond. Sci. Technol. 13 (2000) 1129.
\bibitem{Bordet1} P. Bordet, S. Le~Floch, C. Bougerol-Chaillout, A. Prat, B. Ant\'{e}rion, and M. M\'{e}zouar,
Physica C 341-348 (2000) 577.
\bibitem{Licci} F. Licci, A. Gauzzi, M. Marezio, P. Radaelli, R. Masini, and
C. Chaillout-Bougerol, Physical Review B 58 (1998) 15208.
\bibitem{Karpinski} J. Karpinski, S. M. Kazakov, M. Angst, A. Mironov, M. Mali,
and J.~Roos, Phys. Rev. B 64 (2001) 094518.
\bibitem{Hazen1} R. M. Hazen, C. T. Prewitt, R. J. Angel, N.~L.~Ross, L. W. Finger, C.~G.~Hadidiacos, D. R. Veblen,
P. J. Heaney, P. H. Hor, R. L. Meng, Y.~Y.~Sun, Y. Q. Wang, Y. Y.
Xue, Z. J. Huang, L. Gao, J. Bechtold, and C.~W.~Chu, Phys. Rev.
Lett. 60 (1988) 1174.
\bibitem{Tarascon1} J.~M. Tarascon, Y. Le Page, P.~Barboux, B.~G.~Bagley, L. H. Greene, W.~R.~McKinnon, G.~W.~Hull,
M. Giroud, and D. M. Hwang, Phys. Rev. B 37 (1988) 9382.
\bibitem{Bordet2} P. Bordet, S. M. Loureiro, J. J. Capponi, and P. G. Radaelli,
17th European Crystallographic Meeting (1997).
\bibitem{Toulemonde7} P. Toulemonde, P. Odier, P. Bordet and S.~Le~Floch,
in preparation.

\end{thebibliography}
\end{document}